\documentclass[12pt,preprint]{aastex}

\usepackage{graphicx}
\usepackage{natbib}
\usepackage{rotating}

\newcommand{\be}{\begin{equation}}
\newcommand{\ee}{\end{equation}}

\newcommand{\ba}{\begin{eqnarray}}
\newcommand{\ea}{\end{eqnarray}}

\newcommand{\Swift}{{\it Swift \,}}
\newcommand\etal{\textit{et al.}}
\newcommand\eg{\textit{e.g.\ }}

\newcommand\ie{\textit{i.e.\ }}

\begin{document}
\date{\today}  
\title{Did {\it Swift} measure GRB prompt  emission radii?}
\author{M. LYUTIKOV
}
\affil{University of British Columbia, 6224 Agricultural Road,
Vancouver, BC, V6T 1Z1, Canada and 
  Department of Physics and  Astronomy, University of Rochester,
   Bausch and  Lomb Hall,
    P.O. Box 270171,
     600 Wilson Boulevard,
      Rochester, NY 14627-0171, USA }
\begin{abstract}
The Swift X-Ray Telescope often observes a rapidly decaying
X-ray emission stretching to as long as
  $ t \sim  10^3$ seconds after
  a  conventional prompt phase. This component is  most likely due to
 a prompt emission  viewed at large  observer angles $\theta > 1/\Gamma$, 
where $\theta\sim 0.1$ is a typical
viewing angle of the jet and $\Gamma\geq 100$ is the  Lorentz factor of the flow
during the prompt phase.
  This can be used to estimate the  prompt  emission 
radii, $r_{em} \geq 2  t c /\theta^2 \sim 6 \times 10^{15}$ cm. 
These radii are   much larger than is assumed within a framework of a fireball model.
 Such large emission radii can be reconciled with a fast variability, 
on time scales as short as milliseconds, if the emission 
is beamed in the bulk outflow frame, 
\eg due to a  random relativistic motion of ''fundamental emitters''.
This may also offer a possible explanation for  X-ray flares  observed during early
afterglows.
\end{abstract}


\section{Introduction}
Recently launched
\Swift satellite  \citep{Gehrels} together with  a network of ground based observations
have been providing scientific community with crucial
information on Gamma Ray Bursts (GRBs). Besides the landmark detection of 
afterglows from short  GRBs \citep[\eg][]{Gehr05},  \Swift has 
gathered crucial data on 
developments of GRBs at early  times.
This is especially important since  early observations 
 provide clues to the  properties
of the ejecta, like  its composition, lateral distribution of energy etc.
At late times the energy is mostly transfered to the
forward shock, properties of which can hardly be used to probe the ejecta. 
A number of surprising results related to early afterglows have emerged
\citep[\eg][]{Tagliaferri,Nousek,Chincarini,obrien}:
(i) early, $t \leq 10^3$ s,   rapidly-decaying 
X-ray component, (ii) X-ray flares occurring  at 
  $t \sim 10^2-10^4$ s, (iii) shallower than expected initial decay (or hump) of the afterglow.
These features are common, but the  light curves show a large variety.
  In this letter we discuss the  first  two mentioned effects, \ie rapidly-decaying component 
  and X-ray flares, since both can be related to the prompt emission (as opposed to afterglow)
  and can thus be used to probe the ejecta and the  central engine.

\section{Prompt emission radii}

The initial fast-decaying part of afterglows can be a ''high altitude'' prompt emission,
coming from angles $\theta > 1/\Gamma$ \citep{Kumar00,Barthelmy05},
where  $\theta$ is the angle between the line of sight and the direction from the
center of the explosion towards an emitting point and $\Gamma$ is the Lorentz factor of the outflow.
For a $\delta$-function in time prompt emission pulse, after an initial spike the  observed flux 
should decay as $t^{-(2+\alpha)}$, where $\alpha \approx 0.5  $ is prompt emission's spectral index,
\citep{Fenimore}, roughly consistent with observations.
One also expects that prompt and early afterglow emission 
join smoothly, which seems to be generally observed \citep{obrien}.
[Exceptions, like GRB050219a 
\citep{Tagliaferri},
 may be due to interfering X-ray flares.]

If we accept the interpretation of the fast decaying part as ''high altitude'' prompt emission,
 one can  then
determine  radii of the prompt emission and compare them with model predictions.
 The currently most popular fireball model \cite[\eg][]{Piran04}
 relates radii of emission $r_{em}$ to the variability time scale $\delta t$ 
 of the central source
  $r_{em} \sim
   2 \Gamma_0^2 c \delta t$, where $ \Gamma_0 \sim 100-300$ is the initial Lorentz factor.  
Within the framework of  the fireball model this is also the variability time scale of the 
prompt emission. Observationally, 
prompt emission shows  variability on time scales as short as milliseconds, 
while most power is at a fraction of a second \citep{BeloStern}. Adopting 
$\delta t\sim 0.1
$ s, the prompt emission radius is
 $r_{em} \sim 6 \times 10^{13}$ cm $(\Gamma_0/100)^2$.
If the emission is generated at $r_{em}$ and is coming to observer 
 from large angles,  $\theta > 1/\Gamma$, its delay with respect to the start 
 of the  prompt pulse
is $t \sim (r_{em}/c) \theta^2/2$. If one can estimate $\theta$, then
this can be used to measure $r_{em}$. This can be done
 from  late ''jet breaks'',
giving typically  $\theta \sim 0.1$  \cite[\eg][]{Frail}.
Then, for the  X-ray tail of the prompt   emission extending to $t \sim 1000$ seconds,
 the implied emission radius is
$r_{em} >  6 \times  10^{15}$ cm.
 This is  much larger  
than is assumed in the fireball model.
To make it  consistent with the fireball model and variability on short times scales, the
Lorentz factor of the flow should be huge, $\Gamma_0 \geq 1000$, but this  would imply
that emission is strongly de-boosted, $\Gamma_0 \theta \sim 100  $.
Increasing $\theta$ cannot save the day either since  
 the  required viewing angle would be $\theta \sim 1$, implying a jet moving always from an
 observer.

Along the similar lines of reasoning,
\cite{Lazzati05} estimated prompt emission radii for  a particular case of GRB 050315 
for which a possible jet break is identified
\citep{Vaughan} . Steep decay in that case  is relatively short and 
lasts for 100 s, giving 
$r_{em} > 2.5 \times 10^{14}$ cm. 
Note,
that any observed duration of the steep decay phase provides only  {\it a lower limit}
on the prompt emission  radius since
 the end of the steep decay may be related to
   emergent afterglow emission and not
     to the fact that the edge of the jet becomes visible, see Fig. \ref{GRBafter}.
     On the other hand,
    late jet break  time provides an estimate of the total  opening angle of the jet.
        In any case, GRBs with longer lasting steep decay phase, up to $10^3$ s, provide the
       most severe constraints on the models.

Thus,  the
 interpretation of the fast-decaying initial X-ray light curve
  as prompt emission seen at large angles can hardly be
  inconsistent with
   the  fireball model.
We should then either look for alternative possibilities to produce the  fast-decaying
part of the  X-ray light curve  \citep[\eg][]{mr01}, or consider models that advocate production
of prompt emission at larger radii, see \S \ref{Conc}.

\section{Fast variability from large radii}

If prompt emission is produced at distances $\sim 10^{15}-10^{16}$ cm,
how can a fast variability,
on times scales as short as milliseconds, be 
achieved? One possibility, is that emission is beamed in the outflow frame, for example 
due to a
 relativistic motion of (using pulsar physics parlance)  ''fundamental emitters'' \citep{lb03}.
To prove this point, we 
consider an spherical outflow expanding with a bulk Lorentz factor $\Gamma$ 
with $N$ randomly distributed emitters moving  with respect to the shell rest frame with
 a typical Lorentz factor $\gamma_{T}$.
 Highly boosted emitters, moving towards an  observer,
have a Lorentz factor $
 \gamma \sim 2 \gamma_T \Gamma
$ in the observer frame.
If emission is generated at distances $r_{em}$, the observed variability time scale
can be as short as $ \sim (r_{em}/c) /2  \gamma^2 \approx (r_{em}/c) / 8 ( \gamma_T \Gamma)^2$, so
that   
  modest values of $\gamma_T \sim 5-10 \ll \Gamma \sim 100-300 $
 would suffice to produce a short time scale  variability  from
  large distances $r_{em}\sim 10^{15} - 10^{16} $ cm.

The model should  satisfy  a number of constraints.
 First, the number of sub-jets directed towards an observer from  viewing angles $\theta< 1/\Gamma$ 
 should be larger than unity (in order to produce at least one true prompt emission spike),
 but should not be too large, otherwise prompt emission will be a smooth
 envelope of overlapping spikes. If a typical jet opening angle
 is $\theta_j$, then 
 the number of sub-jets seen ''head-on'' from angles $< 1/\Gamma$ is
\be
n_{prompt} \sim { \pi N  \over (\Gamma \gamma_{T} \theta_j)^2}.
\ee
This should be larger than $1$.

The second constraint that the model should satisfy relates to the
efficiency of energy conversion.
Suppose that the thickness of an outflowing shell in its rest frame is
$ L_{shell} \sim t_s c \Gamma$, where
$t_s $ is a source activity time ($t_s \sim 30-100$ s for long bursts and $t_s \sim 1 $ s
for short  bursts).  Suppose then that fundamental emitters  operate for a 
time $ t_{pulse} = \eta_t L_{shell}$ in the flow frame, where $\eta_t $ is a dimensionless
parameter. During this time the source can tap into energy
contained within  volume $ (c t_{pulse})^3$. The ratio 
of  this volume times the number of emitters to the total volume
of the shell is a measure of efficiency of energy conversion
into radiation:
\be
\eta = { N (c t_{pulse})^3  \over  r_{em}^2 \theta_j ^2 t_s c \Gamma}
\ee
Since tapping  of energy in the volume $(c t_{pulse})^3$ is a definite upper limit on 
conversion efficiency, in the calculations we allow $\eta$ defined above to be slightly 
larger than unity.

 To produce light curves we calculate the intensity of  emission 
 from sub-jets that are randomly located within the shell and moving in random direction
  with random Lorentz factors $1<\gamma_T < \gamma_{T,max} =5 $.
 Each emitter is isotropic in its rest frame and  is active  
  for a random time $0<t'_{em} < \eta_T  t_s c \Gamma
 = t_{pulse,max}$ with $\eta_T  =0.5$.
The observed intensity of  emission  from each sub-jet $\propto \delta^{3+\alpha}$ 
\citep{BlandfordLind},
where $\delta = 1/\gamma (1-\beta \cos \theta_{sj}) $ is a
total Doppler factor  including bulk and  
random motion, $\theta_{sj}$ is an angle between  the line of sight and direction
of the sub-jet motion.
As the burst progresses, larger angles and more of 
 sub-jets producing prompt emission become visible. Most of them will be seen from
  angles  $> 1/\gamma_T$ in the bulk frame, producing a combined smooth curve overlaid
  with spikes.
The average Doppler factor decreases with time $\delta \approx 
 t_s \Gamma / t $ and the average flux decays as $t^{-(2+\alpha)}\approx  t^{-2.5}$ for 
 $\alpha=0.5$.
In Fig. \ref{GRBafter} we  plot an example of a prompt light curve 
in this model. 

\subsection{Lateral dependence of prompt emission}

Variations of the decay rate from the $t^{-(2+\alpha)}$ law may  be used to
probe 
 angular dependence  $L(\theta_{axis})$ of the intensity of the 
prompt emission, where $\theta_{axis}$ is an angle between the axis of the explosion
and an emitting point.  More shallow decays can be due to, \eg,
 a structured jet,  with $L\sim \theta_{axis}^{-2}$ observed outside of some core:
late time emission then is coming from
the more energetic core part.  The effective emission intensity 
  increases approximately as
$\theta ^2 \propto t$,  and will  result in an observed  decay $t^{-(1+\alpha)}$. 
Similarly, if the prompt emission is seen within a core,
late emission comes from less energetic wings, giving in 
case of a  structured jet a flux  $ \propto t^{-(3+\alpha)}$.
Qualitatively, the relativistic internal motion of emitters makes it ''easier'' to see  the
 high altitude emission. 

To show this numerically we parameterize  the {\it number density of emitters}  as
$n(\theta) \propto 1/(\theta^2 + \theta_0^2)$,
where $\theta_0$ is an angular core radius.
[There are, naturally other possible parameterizations, 
e.g. of intensity of each emitter]. The results are presented in Fig. \ref{GRBafterStruct}.

We can also expect deviations from a simple power-law decay due to not exactly spherical
form of the emitting surface. Such distortions are expected due to a  development of the 
Kelvin-Helmholtz   instability during an accelerating phase of the outflow.
They won't be erased during the  coasting stage due to causal disconnection of the flow
separated by angles $> 1/\Gamma$. 
Additional complications
may come from the way the data analysis is performed, \eg through a choice of 
initial time trigger (\cite{Zhang05}, see also \cite{Lazzati05}).

\section{Origin  of X-ray flares}

Early X-ray light curves show  complex behavior with  flares and
frequent changes in a temporal slope \citep[\eg][]{obrien}. Flares
    show very short rise and fall times, much shorter than
     observation time after the on-set of a GRB, while the 
     underlying afterglow has the same  behavior before and
              after the flare \citep{Burrows}
	      (though there are exceptions).
	      Both of these observations
     argue against a 
     physical process in the forward shock.
In addition, there is a hardening of the  spectrum during  X-ray flares \citep{Burrows}.

In the present model we interpret  X-ray flares as been due to  sub-jets located at 
large viewing  angles, $\theta > 1/\Gamma$, but
 directed towards an observer.
 Randomly located, narrow spikes are clearly seen in the model light curves,
  Figs. \ref{GRBafter}-\ref{GRBafterStruct}.
In addition, as  the flares are  less de-boosted than the average high altitude  outflow, 
they will have a harder spectrum, as observed.

\section{Discussion}
\label{Conc}

In this letter we first point out 
 that the interpretation of  the initial fast-decaying part of the X-ray GRB
light curves as a prompt emission seen at large angles,  and a generic estimate of 
jets' opening angle allows a measurement of the
 radius of prompt emission, which turns out to be relatively large, 
 $> 10^{15}$ cm. 
 On basic grounds, $\gamma$-ray  emission should be generated before the deceleration radius 
 $r_{dec} \sim \left( { E_{iso} \over 4 \pi \rho c^2 \Gamma_{dec}^2} \right)^{1/3}\sim 10^{16}-
 10^{17}$ cm, 
 when most energy of the outflow is given to the surrounding medium
 (here $ E_{iso}$ is isotropic equivalent energy,  $ \rho$ is density of external medium,
 $\Gamma_{dec}$ is  Lorentz factor at $r_{dec}$).
 \footnote{Note, that $r_{dec}$ defined above is {\it independent} of ejecta content, contrary to 
 the claim in
 \cite{zk05}, see \cite{LZK}.}
 The inferred emission radius is within this limit. 

The estimate of the emission radius is very simple, and, in some sense, generic.
It can hardly be consistent with the fireball model, unless extreme assumptions are made about
the parameters (\eg very large Lorentz factor).
         On the other hand, 
  there are  alternative models (\eg the 
  electromagnetic model  \citep{l05}, see also \cite{thompson})
 that  place prompt emission radii  at large distances, just before the
 deceleration radius
  $r_{dec} $.

Secondly, we show how   models  placing emission at large radii
 may be able to reproduce a short time scale variability of the prompt emission
 and explain later  X-ray flares. This can be achieved if  the prompt emission is beamed
 in the rest frame of the outflow, which may be due to  an internal relativistic motion
 of ''fundamental emitters''.

What can produce a relativistic motion  in the bulk frame? 
It can be due, for example, to a  relativistic Burgers-type turbulence 
(a collection  of randomly directed shock waves). It is not clear how such turbulence may be generated.
Alternatively,
 relativistic internal sub-jets
can  result  from   reconnection occurring in highly magnetized  plasma  
with $\sigma \gg 1$,
where
$\sigma$  is a plasma magnetization parameter \citep{KC84}.
In this case
 the  matter outflowing from a reconnection layer  reaches
relativistic speeds with $\gamma_{out} \sim \sigma$ \cite[]{lu03}.
Internal synchrotron emission by such jets,
or 
Compton scattering of ambient photons, will  be strongly beamed in the frame of the outflow.
Note, that {\it this model does not require
late engine activity } to produce flares. 

\begin{figure}
\includegraphics[width=0.95\linewidth]{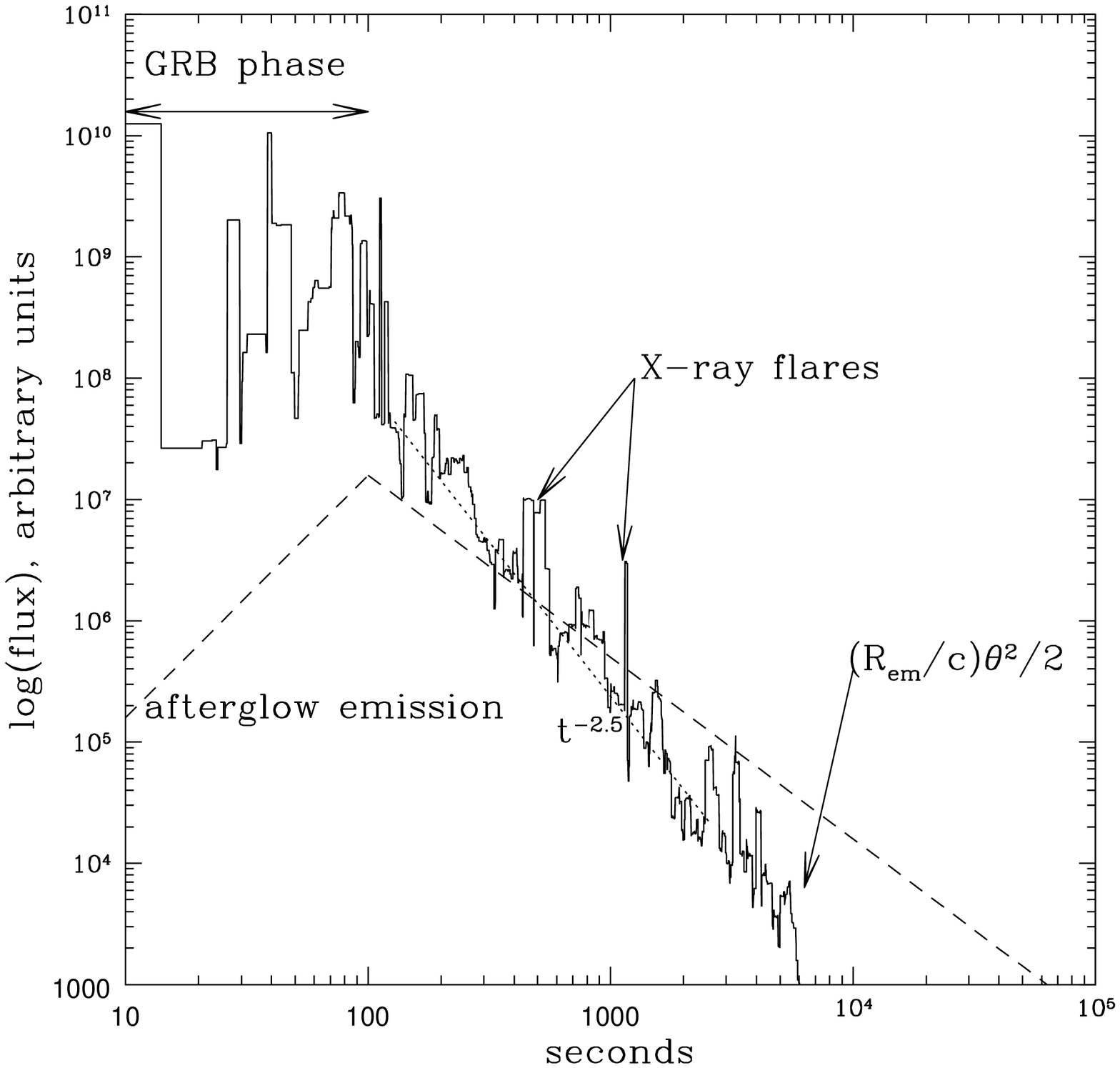}
\caption{Prompt emission produced by emitters  moving  randomly  in the bulk frame.
Emission is generated within a shell of thickness $t_s c = 3 \times 10^{12}$ cm 
in observer frame,
moving with $\Gamma =100$  at distance $r_{em} = \Gamma^2 t_s c $ by randomly distributed
sub-jets with random orientation
moving with random Lorentz factors $1<\gamma_T < \gamma_{T,max} =5 $.
Each emitter is active for random time $0<t'_{em} < 0.5 t_s c \Gamma = t_{pulse,max}$ 
in its rest frame.
Dotted line: average intensity $\propto t^{-(2+\alpha)} =t^{-2.5}$.
Dashed line: expected afterglow signal rising $\propto t^2$, peaking
at $\sim 100$ s and falling off $\propto t^{-1.5}$ with arbitrary normalization.
Homogeneous jet centered on an observer with opening  angle $\theta =0.1$,
dimensionless parameters 
 are $n_{prompt} =1.2$ and $\eta=1.6$.
}
\label{GRBafter}
\end{figure}
\begin{figure}
\includegraphics[width=0.95\linewidth]{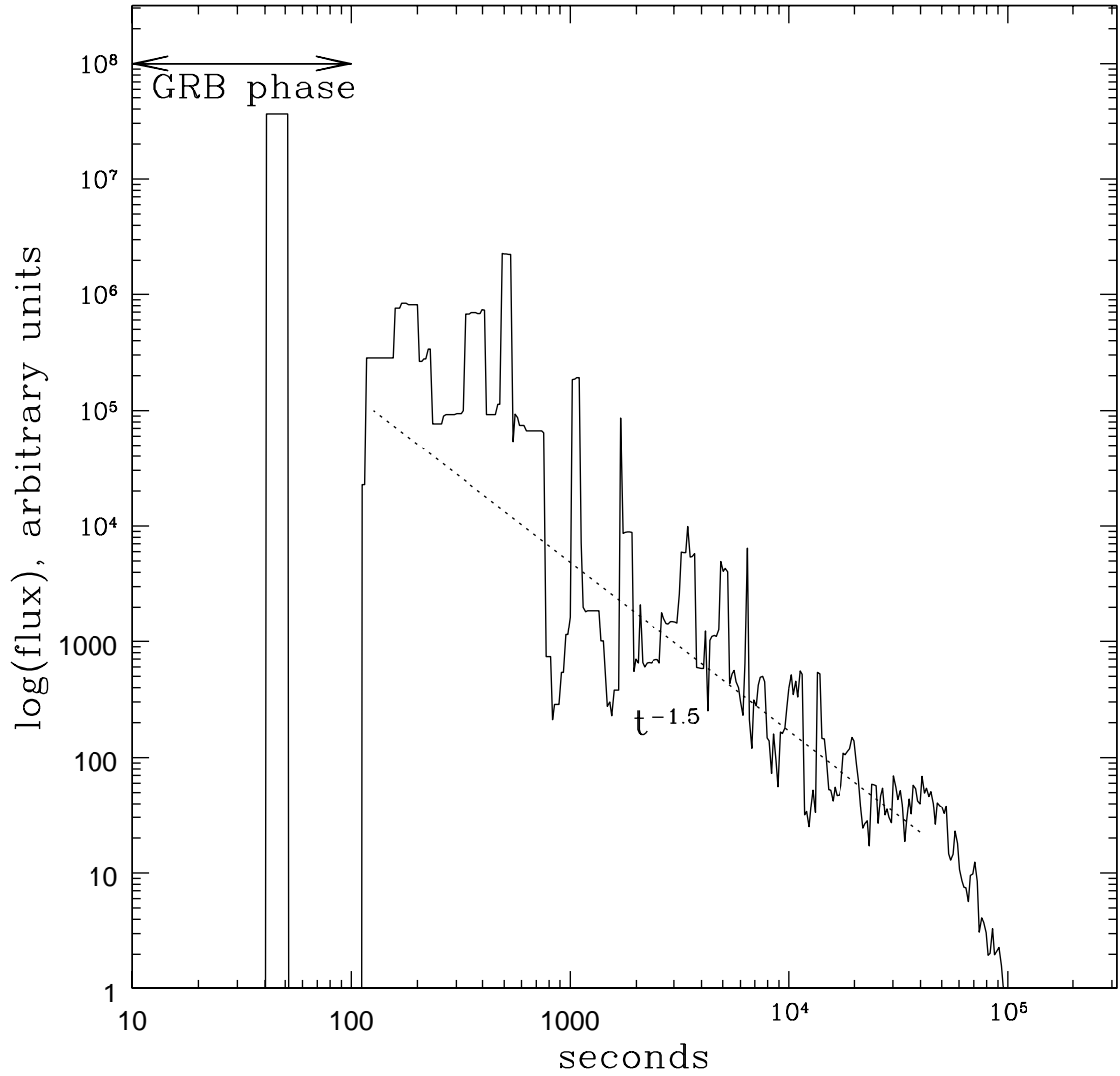}
\caption{Same as Fig. \ref{GRBafter} 
for structured jet with number density of sub-jets $\propto 1/(\theta^2 + \theta_0^2)$,
core size $\theta_0 = \pi/100$, observer angle $\theta_{ob} = \pi/10$,
dimensionless parameters  are 
$n_{prompt} =1.26$, $\eta = 1.6$ (calculated with $\theta=\theta_{ob}$). 
}
\label{GRBafterStruct}
\end{figure}

One of the main observational complications is that at observer times larger than  the
conventional prompt phase, the X-ray light curve is a sum of the tail of the prompt emission,
coming presumably from internal dissipation in the ejecta, and the forward shock emission. 
It is not obvious how to separate the two  components.
For example, GRBs which do not show a fast initial decay may be 
dominated by the forward shock emission from early on
 \citep{obrien}.
 This uncertainty also affects estimates of the emission radius
since the  end of the steep decay may be related to
 the emergent afterglow emission and not to the jet opening  angle (or observer's  
 angle, in case of
 a structured jet), see Fig. \ref{GRBafter}.
 Another complication is that at these intermediate times,
 $10^3 \leq t  \leq 10^4$ s,  even the 
forward shock emission  itself  often
 does not conform to the standard afterglow
models, showing  flatter than expected 
profiles \citep[\eg][]{Nousek}.

A consequence of the model is that {\it some} short GRBs may be just a single
spike directed towards an observer of a long GRBs. In our model
the shorter spikes are  highly beamed, less frequent and produce harder emission. 
 This can apply only to {\it some} short GRBs since as a class
they are well established to have different origin than  long GRBs
(from non-observation of a supernova signature and coming from a distinctly different
host galaxy population). 


\end{document}